\documentclass[a4paper]{jpconf}
\usepackage{graphicx}
\usepackage{amsmath}
\usepackage{amsfonts, amsmath, amsthm, amssymb}
\usepackage[font=small,labelfont=bf]{caption}

\begin{document}
\title{Collective neutrino oscillations in moving and polarized matter}

\author{Zekun Chen$^{1}$, Konstantin Kouzakov$^{1}$, Yu-Feng Li$^{2,3}$, Vadim Shakhov$^{1}$, Konstantin Stankevich$^{1,4}$ and Alexander Studenikin$^{1,5}$}

\address{$^{1}$ Faculty of Physics, Lomonosov Moscow State University, Moscow 119991, Russia}
\address{$^{2}$ Institute of High Energy Physics, Beijing, China}
\address{$^{3}$School of Physical Sciences, University of Chinese Academy of Sciences, Beijing 100049, China}
\address{$^{4}$ Kotel’nikov Institute of Radio Engineering and Electronics RAS, 125009, Moscow, Russia}
\address{$^{5}$ Joint Institute for Nuclear Research, Dubna 141980, Moscow Region, Russia}

\ead{liyufeng@ihep.ac.cn, studenik@srd.sinp.msu.ru}

\begin{abstract}
We consider neutrino evolution master equations in dense moving and polarized matter consisted of electrons, neutrons, protons and neutrinos. We also take into account the neutrino magnetic moment interaction with a magnetic field. We point out the mechanisms responsible for the neutrino spin precession and provide the expressions for the corresponding interaction Hamiltonians that should be taken into account in theoretical treatments of collective neutrino oscillations.  
\end{abstract}

\section{Introduction}
Future large volume detectors, such as JUNO, Hyper-Kamiokande and DUNE, open a new era in studies of supernovae and presupernovae neutrinos \cite{Totani:1997vj,Li:2020gaz}. Astrophysical neutrinos would give an important information for the explosion mechanism of collapse-driven supernovae. The present our study is dedicated to the neutrino evolution in astrophysical environment.
In most papers dedicated to the collective neutrino oscillations, one usually considers transitions between flavour states without change of helicity. The authors of \cite{deGouvea:2012hg,deGouvea:2013zp} for the first time studied collective neutrino oscillation accounting for transitions between states with different helicity engendered by interaction of the neutrino magnetic moment with a magnetic field. Later, the problem of spin oscillations in collective effects was studied by different scientific groups (see, for example, \cite{Yuan:2021exm,Abbar:2020ggq} and references therein).
In \cite{Chatelain:2016xva} the effect of spin oscillations engendered by transversally moving matter was included in collective neutrino oscillation.
In \cite{Chatelain:2017yxx,Stapleford:2016jgz} collective neutrino oscillations were studied in the case of neutrino nonstandard interactions.
Here below, we summarize all effects that can lead to neutrino spin precession in collective neutrino oscillations in the presence of moving and polarized matter.

\section{Neutrino evolution in astrophysical environment}
Consider two flavor neutrinos with two possible helicities $\nu_f = (\nu^-_e, \nu^-_x, \nu^+_e, \nu^+_x)$, where $\nu_x$ stands for $\nu_\mu$ or $\nu_\tau$. Then the neutrino system is described by the density matrix (we follow the formalism developed in \cite{deGouvea:2012hg,deGouvea:2013zp})

\begin{equation}
\rho =
\left(
\begin{matrix}
\rho_\nu & X \\
X^\dagger & \rho_{\bar \nu}
\end{matrix}
\right)
,
\end{equation}
where $\rho$ and $\rho_{\bar \nu}$ are the usual $2 \times 2$ flavour density matrices. In the case of Dirac neutrinos the matrices describe active and sterile neutrinos, respectively. In the case of Majorana neutrinos $\rho_{\bar \nu}$ describes antineutrinos. The external magnetic field and the transversally moving matter leads to the coupling between $\rho$ and $\rho_{\bar \nu}$ (see \cite{Pustoshny:2018jxb} and references therein) and, therefore, we should also consider non-diagonal matrices $X$

\begin{equation}
X =
\left(
\begin{matrix}
\rho_{\nu_e \bar \nu_e} & \rho_{\nu_e \bar \nu_x} \\
\rho_{\nu_x \bar \nu_e} & \rho_{\nu_x \bar \nu_x}
\end{matrix}
\right)
.
\end{equation}

The evolution of $\rho$ is governed by Liouville-von Neumann master equation

\begin{equation}
i \dfrac{d \rho}{d t} = [H, \rho]
,
\end{equation}
where $H$ is the Hamiltonian that describes the external environment. It consists of five parts

\begin{equation}
H = H_{vac} + H_B + H_{mat} + H^f_\zeta  +H_{\nu\nu} 
,
\end{equation}
where $H_{vac}$ is the vacuum Hamiltonian, $H_B$ is the Hamiltonian that accounts for the magnetic field, $H_{mat}$ is the matter potential for electrons and neutrons  moving in an arbitrary direction, $H^f_\zeta$ is the Hamitonian that accounts for the matter polarization. The neutrino-neutrino interaction potential, $H_{\nu\nu}$, is

\begin{equation}
H_{\nu\nu} = \sqrt{2} G_F n_\nu \int dE \left[ G^\dagger (\rho(E) - \rho(E)^{c*}) G + \dfrac 1 2 G^\dagger \Tr \left[ (\rho(E) - \rho(E)^{c*}) G \right] \right]
,
\end{equation}
with $G_F$ being the Fermi coupling constant and $n_\nu$ being a neutrino density profile. The density matrix $\rho^c$ is define in the same way as in \cite{Abbar:2020ggq}

\begin{equation}
\rho^c =
\left(
\begin{matrix}
\rho_{\bar \nu} & X^* \\
X^T & \rho_\nu
\end{matrix}
\right)
.
\end{equation}

Dimensionless matrix $G$ is

\begin{equation}
G =
\left(
\begin{matrix}
+1 & 0 & 0 & 0\\
0 & +1 & 0 & 0\\
0 & 0 & -1 & 0\\
0 & 0 & 0 & -1
\end{matrix}
\right)
.
\end{equation}

The Hamiltonian that accounts for the neutrino magnetic moment interaction with parallel and perpendicular components of the magnetic field $B = B_{||} + B_{\perp}$ has the following form

\begin{equation}
H^D_B=
\begin{pmatrix}
\left(\frac{\mu}{\gamma}\right)_{ee}B_{||} & \left(\frac{\mu}{\gamma}\right)_{ex}B_{||} & -\mu_{ee}B_{\perp}e^{i\phi} & -\mu_{ex}B_{\perp}e^{i\phi} \\

\left(\frac{\mu}{\gamma}\right)_{ex}B_{||} & \left(\frac{\mu}{\gamma}\right)_{xx}B_{||} & -\mu_{ex}B_{\perp}e^{i\phi} & -\mu_{xx}B_{\perp}e^{i\phi} \\

-\mu_{ee}B_{\perp}e^{-i\phi} & -\mu_{ex}B_{\perp}e^{-i\phi} & -\left(\frac{\mu}{\gamma}\right)_{ee}B_{||} & -\left(\frac{\mu}{\gamma}\right)_{ex}B_{||} \\

-\mu_{ex}B_{\perp}e^{-i\phi} & -\mu_{xx}B_{\perp}e^{-i\phi} & -\left(\frac{\mu}{\gamma}\right)_{ex}B_{||} & -\left(\frac{\mu}{\gamma}\right)_{xx}B_{||}
\end{pmatrix}
,
\end{equation}
where $\phi$ is the angle between $\textbf{v}_{\perp}$ and $\textbf{B}_{\perp}$. 

Magnetic moments in flavour basis $\mu_{\alpha \beta}$ are expressed through magnetic moments $\mu_{ij}$ in mass states

\begin{equation}
\begin{aligned}
&\mu_{ee} = \mu_{11}\cos^2\theta + \mu_{22}\sin^2\theta + \mu_{12}\sin 2\theta ,\\
&\mu_{ex} = \mu_{12}\cos2\theta + \frac12\left(\mu_{22} - \mu_{11}\right)\sin 2\theta ,\\
&\mu_{xx} = \mu_{11}\sin^2\theta + \mu_{22}\cos^2\theta - \mu_{12}\sin 2\theta ,
\end{aligned}
\end{equation}
for neutrino interaction with $B_\perp$ and

\begin{equation}
\begin{aligned}
&\left(\frac{\mu}{\gamma}\right)_{ee} = \frac{\mu_{11}}{\gamma_{11}}\cos^2\theta + \frac{\mu_{22}}{\gamma_{22}}\sin^2\theta + \frac{\mu_{12}}{\gamma_{12}}\sin 2\theta ,\\
&\left(\frac{\mu}{\gamma}\right)_{ex} = \frac{\mu_{12}}{\gamma_{12}}\cos2\theta + \frac12\left(\frac{\mu_{22}}{\gamma_{22}} - \frac{\mu_{11}}{\gamma_{11}}\right)\sin 2\theta ,\\
&\left(\frac{\mu}{\gamma}\right)_{xx} = \frac{\mu_{11}}{\gamma_{11}}\sin^2\theta + \frac{\mu_{22}}{\gamma_{22}}\cos^2\theta - \frac{\mu_{12}}{\gamma_{12}}\sin 2\theta ,\\
\end{aligned}
\end{equation}
for neutrino interaction with $B_{||}$. Here we have introduced the following notifications

\begin{equation}
\gamma_{\alpha}^{-1}=\frac{m_{\alpha}}{E_{\alpha}} , \ \ \ \ \
\gamma_{\alpha\beta}^{-1}=\frac12\left(\gamma_{\alpha}^{-1} + \gamma_{\beta}^{-1}\right) ,\ \ \ \ \
\tilde \gamma_{\alpha\beta}^{-1}=\frac12\left(\gamma_{\alpha}^{-1} - \gamma_{\beta}^{-1}\right) .
\end{equation}

In the case of neutrino is a Majorana particle the diagonal magnetic moments are zero in both flavour and mass basis. Therefore, the Hamiltonian for Majorana neutrino in the magnetic field is expressed as

\begin{equation}
H^M_B = i\mu\cos2\theta
\begin{pmatrix}
0 & \frac{1}{\gamma_{12}}B_{||} & 0 & -B_{\perp}e^{i\phi} \\

-\frac{1}{\gamma_{12}}B_{||} & 0 & B_{\perp}e^{i\phi} & 0 \\

0 & -B_{\perp}e^{-i\phi} & 0 & -\frac{1}{\gamma_{12}}B_{||} \\

B_{\perp}e^{-i\phi} & 0 & \frac{1}{\gamma_{12}}B_{||} & 0
\end{pmatrix}
.
\end{equation}

The Hamiltonian that accounts for the electron and neutron matter is
\begin{equation}
H^D_{mat}= \frac{G_F}{2\sqrt{2}}
\begin{pmatrix}
2(2 n_e -  n_n)\left(1-v_{||}\right)  & 0 & (2 n_e -  n_n)  v_{\perp}\left(\frac{\eta}{\gamma}\right)_{ee}  & (2 n_e -  n_n) v_{\perp} \left(\frac{\eta}{\gamma}\right)_{ex} \\

0 & - 2n_n\left(1-v_{||}\right)  &  - n_n v_{\perp} \left(\frac{\eta}{\gamma}\right)_{ex} & - n_n v_{\perp} \left(\frac{\eta}{\gamma}\right)_{xx}  \\

(2 n_e -  n_n) v_{\perp} \left(\frac{\eta}{\gamma}\right)_{ee}  & - n_n v_{\perp} \left(\frac{\eta}{\gamma}\right)_{ex} & 0 & 0 \\

(2 n_e -  n_n) v_{\perp} \left(\frac{\eta}{\gamma}\right)_{ex} & - n_n v_{\perp}  \left(\frac{\eta}{\gamma}\right)_{xx} & 0 & 0
\end{pmatrix}
,
\end{equation}
where $n_n$ and $n_e$ are the neutron and electron density profiles $n_{n(e)}=\frac{n^{0}_{n(e)}}{\sqrt{1-v^{2}}}$, the electron and neutron velocity is $v = v_{||} + v_{\perp}$ and

\begin{equation}
\left(\frac{\eta}{\gamma}\right)_{ee} = \frac{\cos^2\theta}{\gamma_{11}} + \frac{\sin^2\theta}{\gamma_{22}} ,\ \ \ \ \
\left(\frac{\eta}{\gamma}\right)_{xx} = \frac{\sin^2\theta}{\gamma_{11}} + \frac{\cos^2\theta}{\gamma_{22}} ,\ \ \ \ \
\left(\frac{\eta}{\gamma}\right)_{ex} = \frac{\sin2\theta}{\tilde \gamma_{21}}
.
\end{equation}

The matter polarization is taken into account by the following Hamiltonian in the flavour basis

\begin{equation}\label{H_v}
H^f_\zeta=n_e \frac{G_F}{2\sqrt{2}}
\begin{pmatrix}
0 & (\frac{\eta}{\gamma})_{ee}\zeta_{\perp} & 0 &(\frac{\eta}{\gamma})_{e\mu}\zeta_{\perp}\\
(\frac{\eta}{\gamma})_{ee}\zeta_{\perp} & 2\zeta_{\parallel} & (\frac{\eta}{\gamma})_{e\mu}\zeta_{\perp} & 0\\
0 & (\frac{\eta}{\gamma})_{e\mu}\zeta_{\perp} & 0 &(\frac{\eta}{\gamma})_{\mu \mu}\zeta_{\perp}\\
(\frac{\eta}{\gamma})_{e\mu}\zeta_{\perp} & 0 &(\frac{\eta}{\gamma})_{\mu\mu}\zeta_{\perp} & 2\zeta_{\parallel}\\
\end{pmatrix},
\end{equation}
where $\zeta_{||}$ and $\zeta_{\perp}$ are the longitudinal and transversal polarization of the matter in respect to the direction of neutrino propagation.

\section{Summary}
We have considered the mechanisms of neutrino spin precession in collective neutrino oscillations. Besides the neutrino magnetic moment interaction with a magnetic field the discussed mechanisms include the neutrino interaction with both transversally moving and transversally polarized matter. We have formulated the corresponding effective interaction Hamiltonian for each mechanism. The developed formalism provides a general basis for studying the neutrino spin precession effects in collective neutrino oscillations.

\section*{Acknowledgments}
This work was supported by the joint project of the National Natural Science Foundation
of China (No.~12011530068), and the Russian Foundation for Basic Research (No.~20-52-53022-
GFEN-a), by the National Natural Science Foundation of China under Grant No. 11835013, and
No. 12075255, by Beĳing Natural Science Foundation under Grant No. 1192019,
by the Interdisciplinary Scientific and Educational School of Moscow University “Fundamental and Applied Space Research”. The work of KS is also supported by the RFBR under grant No.~20-32-90107 and by the “BASIS” Foundation No.~20-2-2-3-1.

\end{document}